\documentclass[10pt,journal,compsoc]{IEEEtran}

\usepackage[utf8]{inputenc}
\usepackage[switch]{lineno}

\usepackage{listings, xcolor}

\definecolor{verylightgray}{rgb}{.97,.97,.97}

\lstdefinelanguage{Solidity}{
	keywords=[1]{anonymous, assembly, assert, balance, break, call, callcode, case, catch, class, constant, continue, constructor, contract, debugger, default, delegatecall, delete, do, else, emit, event, experimental, export, external, false, finally, for, function, gas, if, implements, import, in, indexed, instanceof, interface, internal, is, length, library, log0, log1, log2, log3, log4, memory, modifier, new, payable, pragma, private, protected, public, pure, push, require, return, returns, revert, selfdestruct, send, solidity, storage, struct, suicide, super, switch, then, this, throw, transfer, true, try, typeof, using, value, view, while, with, addmod, ecrecover, keccak256, mulmod, ripemd160, sha256, sha3}, 
	keywordstyle=[1]\lst@ifdisplaystyle\color{blue}\bfseries \fi,
	keywords=[2]{address, bool, byte, bytes, bytes1, bytes2, bytes3, bytes4, bytes5, bytes6, bytes7, bytes8, bytes9, bytes10, bytes11, bytes12, bytes13, bytes14, bytes15, bytes16, bytes17, bytes18, bytes19, bytes20, bytes21, bytes22, bytes23, bytes24, bytes25, bytes26, bytes27, bytes28, bytes29, bytes30, bytes31, bytes32, enum, int, int8, int16, int24, int32, int40, int48, int56, int64, int72, int80, int88, int96, int104, int112, int120, int128, int136, int144, int152, int160, int168, int176, int184, int192, int200, int208, int216, int224, int232, int240, int248, int256, mapping, string, uint, uint8, uint16, uint24, uint32, uint40, uint48, uint56, uint64, uint72, uint80, uint88, uint96, uint104, uint112, uint120, uint128, uint136, uint144, uint152, uint160, uint168, uint176, uint184, uint192, uint200, uint208, uint216, uint224, uint232, uint240, uint248, uint256, var, void, ether, finney, szabo, wei, days, hours, minutes, seconds, weeks, years},	
	keywordstyle=[2]\lst@ifdisplaystyle\color{teal}\bfseries \fi,
	keywords=[3]{block, blockhash, coinbase, difficulty, gaslimit, number, timestamp, msg, data, gas, sender, sig, value, now, tx, gasprice, origin},	
	keywordstyle=[3]\lst@ifdisplaystyle\color{violet}\bfseries \fi,
	identifierstyle=\color{black},
	sensitive=false,
	comment=[l]{//},
	morecomment=[s]{/*}{*/},
	commentstyle=\color{gray}\ttfamily,
	stringstyle=\color{red}\ttfamily,
	morestring=[b]',
	morestring=[b]"
}

\usepackage{hyperref}
\usepackage{graphicx}
\usepackage{multirow}
\usepackage{pifont}
\newcommand*\OK{\ding{51}}
\newcommand*\NOK{\ding{55}}

\begin{document}


\title{Testing Smart Contracts Gets Smarter}

\author{\IEEEauthorblockN{Erfan Andesta \IEEEauthorrefmark{1},
		Fathiyeh Faghih\IEEEauthorrefmark{2} and Mahdi Fooladgar\IEEEauthorrefmark{3} \\}
	\IEEEauthorblockA{College of Engineering, Department of Electrical and Computer Engineering \\ University of Tehran, Tehran, Iran \\
		Email: \IEEEauthorrefmark{1}andesta.erfan@ut.ac.ir,
		\IEEEauthorrefmark{2}f.faghih@ut.ac.ir,
		\IEEEauthorrefmark{3}m.fooladgar@ut.ac.ir}}

%
	\IEEEtitleabstractindextext{%
\begin{abstract}
Smart contracts are immutable, verifiable, and autonomous pieces of code that can be deployed and ran on  blockchain networks like Ethereum. Due to the immutability nature of blockchain, no change is possible on a deployed smart contract or a verified transaction. On the other hand, there are millions of dollars carried by smart contracts in Ethereum blockchain, and hence, a faulty smart contract can lead to a huge monetary loss. Therefore, it is important for smart contract developers to fully test and check the correctness of their code before deploying it on the blockchain. In this paper, we propose a testing mechanism for smart contracts in Solidity language, based on mutation testing. We analyzed a comprehensive list of known bugs in Solidity smart contracts, and designed 10 classes of mutation operators inspired by the real faults. Our experimental results show that our proposed mutation operators can regenerate 10 of 15 famous faulty smart contracts, which have resulted in millions of dollars loss. The results show the effectiveness of our proposed mutation operators in detecting real faults in Solidity smart contracts. We have also extended {\em Universal Mutator } tool with our mutation operators, so that it can automatically generate mutants for smart contracts written in Solidity.
\end{abstract}

\markboth{Journal of IEEE Software}%
{Shell \MakeLowercase{\textit{et al.}}: Testing Smart Contracts Gets Smarter}


		\begin{IEEEkeywords}
Blockchain, Smart Contracts, Security Testing, Mutation Testing, Solidity Language
	\end{IEEEkeywords}}

	\maketitle


\section{Introduction}\label{sec:introduction}
Blockchain is an immutable distributed ledger, where users can trigger transactions  by creating a wallet~\cite{nakamoto2008bitcoin}. 
In blockchain, there is no central database, and the full history of transactions is stored by all network nodes. Adding new transactions is only devoted to {\em miners}, who try to produce new blocks out of received transactions to earn rewards.
In late 2013, Buterin et al. published the Ethereum white paper~\cite{buterin2014ethereum}, where they introduced Ethereum as a global, open-source platform for decentralized applications (known as {\em smart contracts}) based on blockchain.
Smart contracts are immutable pieces of code, stored and executed autonomously by the Ethereum miners. They can hold \textit{Ether} or new defined assets (ERC20 tokens), and include rules for sending (or dispensing) these assets. Several smart contracts have been published on the Ethereum network for different applications, including security tokens, voting, gambling and lottery, property ownership, stocktaking, etc. Millions of dollars are carried by smart contracts, and hence, any security vulnerability in these smart contracts can lead to huge monetary losses. As an example, we can mention the famous DAO attack~\cite{daoattack}, which resulted in loss of 150 million dollars. Due to the immutable nature of smart contracts, their correctness is significant. There are more than 34000 vulnerable smart contracts on the Ethereum blockchain, carrying about 4905 \textit{Ether}~\cite{nikolic2018finding}, which shows the need for effective analysis techniques for them. Therefore, guaranteeing the correctness of smart contracts have recently attracted researchers in software engineering and distributed computing.  


There have been several attempts to analyze the correctness of smart contracts. Most of these works focus on static analysis of smart contracts based on the known bug patterns~\cite{luu2016making,kalra2018zeus}. Static analysis is a strong tool for evaluating software quality, however, it has its own limitations. 
First of all this technique can only find the bugs which are matched with the known patterns, and any other bug cannot be detected. Moreover, since the code is not executed in this technique, there may be many false positives in the reported bugs. In other words, some reported bugs may correspond to the paths of the code which are not possible to execute. Therefore, static analysis techniques need to be strengthened with dynamic analysis methods including testing. One of the main challenges in testing is test design. More specifically, from the large set of possible scenarios, the question is how to select a sufficiently small subset of test cases that are more likely to find the potential bugs. Several techniques have been introduced in the test community for designing effective test cases. It is crucial to have a test design technique, considering the specific features of smart contracts, and the bugs resulting from the distributed nature of their running environment.
Our idea is to use mutation testing approach for smart contract testing. Mutation testing is known as the strongest technique for test design. There are several mutation operators designed for different programming languages. In this paper, our goal is to propose a set of mutation operators specifically designed for the Solidity programming language, considering the known real bugs made by smart contract developers. There is one paper with similar idea~\cite{wu2019mutation}, where 15 mutation operators are proposed {\em based on Solidity documentation}. Note that the power of mutation testing is very much dependent on its mutation operators, and we believe the operators that can mimic the real bugs can select more effective test cases. We did an extensive study to identify the known bugs in the scope of smart contracts using many academic papers, open-source smart contracts (and their open or closed issues) on Github~\cite{github}, articles in the websites like Medium~\cite{medium}, and Q\&A websites like Stack-Exchange~\cite{exchange}. 
We used this list of real world known bugs to design a set of mutation operators. We also proposed a set of mutation operators for Solidity specific features, including constructor, modifiers, methods for transaction call, and also self-destruct.
\\
Our contributions in this work are as follows: 
\begin{itemize}
  \item We propose a set of mutation operators for Solidity inspired by the real faults in smart contracts written in this language.
  \item We evaluated the effectiveness of our mutation operators by applying them to the fixed version of famous buggy smart contracts, and finding the percentage of real bugs generated by our operators. 
  \item We extended the {\em Universal Mutator}~\cite{universal} tool by our mutation rules to generate mutants for smart contracts automatically.
\end{itemize}

This paper is organized as follows. In Section~\ref{ssec:Preliminaries}, we briefly discuss the preliminary concepts including smart contracts and mutation testing. In Section~\ref{sec: Mutation Testing of Smart Contracts}, we present our mutation operators with details about the intuition behind them. In Section \ref{Evaluation}, we discuss our techniques for evaluating the effectiveness of our mutation operators. We briefly mention the related works in Section~\ref{ssec:Related Work}, and concluding remarks are presented in Section~\ref{Conclusion}.
\section{Preliminaries}\label{ssec:Preliminaries}
In this section, we briefly discuss the preliminary concepts used in this work, including the notion of smart contract, as well as mutation testing concepts.

\subsection{Smart Contracts}\label{ssec:Smart Contracts}
As we discussed earlier, Ethereum provides a Turing-complete programming language named Solidity for developing smart contracts. Each smart contract is an Ethereum account that holds a piece of code (i.e. the source code for the smart contract), has some private storage, and also can hold some money in \textit{Ethers} (the Ethereum currency). Thus, there are two types of accounts in Ethereum:
 \begin{itemize}
	\item External accounts, which are owned by Ethereum users and are like simple bank accounts, and
	\item Smart contract accounts
\end{itemize} 
 
Both users and smart contracts can initiate transactions for transferring currency or running a smart contract, which will change the state of the transaction-based Ethereum Virtual Machine ({\em EVM}). The execution of smart contracts and verifying the validity of transactions are devoted to a set of special network nodes, called {\em miners}, which get fee from the transactions' initiators for doing the computations.

In the Ethereum network, smart contracts can be written it different languages such as JavaScript, C++, and Solidity. To deploy a smart contract on Ethereum, it should be compiled to EVM byte-code and stored on the Ethereum network. Then, the EVM allocates a limited storage for the compiled contract and assigns an address to both compiled smart contract and the allocated storage. Any Ethereum user can later use the assigned address to trigger any function from the compiled contract. 
To submit a transaction, either for transferring money or triggering a function from a smart contract, a user should pay some fee in {\em Ether}, known as {\em gas}. Paying the gas prevents the Ethereum network from being attacked by wasteful tasks (or getting stuck in an infinite loop), and also is an incentive for the miners to run the smart contract or verify the transaction. In the Ethereum network, the gas amount is related to the complexity of the code, meaning that an Ethereum user needs to pay more gas to run complex smart contracts. In a case that a user do not pay sufficient gas value, the EVM raises an exception and all changes caused by the transaction will revert. 
\subsection{Mutation Testing}\label{ssec:Mutation Testing}
Mutation Testing is one of the strongest techniques for test design. The idea behind mutation testing is changing the source code of a program with tiny syntactic changes, called \textit{mutation operators}. The mutation operators are language specific changes that are designed considering the features of the language, or the common faults made by software developers.
For example, a mutation operator may change the $+$ mathematical operator to $-$, or remove negation from a logical condition. 
To test a program, we should apply all mutation operators to the source code of the program (one after another), which results in a set of modified programs, called \textit{mutants}. 
To select a set of test cases from  a pool of randomly generated scenarios, they are run on the program and the mutants. 
If the output of running the test case on the program is different from the output of running it on a mutant, the mutant is said to be {\em killed}, and that test case is selected. {\em Mutation score} for a set of test cases corresponds to the percentage of mutants killed by these scenarios, and is a metric for evaluating the effectiveness of test cases. 

The philosophy of mutation testing is based on two hypotheses: the competent programmer hypothesis (CPH) \cite{budd1979mutation},\cite{demillo1978hints}, and the coupling effect hypothesis \cite{demillo1978hints}.
CPH states that the competition among the  developers makes them develop a code that is really close to the final correct program. This indicates that although there may be faults in a program, but most of them are minor, and can be fixed by minor changes in the program. Therefore, mutation operators are designed  to add tiny changes in the program.
The Coupling Effect states that a test case identifying the difference between a small changing  mutant and the original programs, can identify more complicated faults as well.

\section{Mutation Testing of Smart Contracts}\label{sec: Mutation Testing of Smart Contracts}
In this research, our goal is to propose a systematic approach for selecting effective test cases  for smart contracts based on mutation.  The strength of this method is very much dependent to its mutation operators. Mutation operators are usually inspired by the common faults in a language or domain. In other words, they usually mimic the common bugs of the programs written in a language. Different mutation operators may be applicable for many programming languages, while some of them may be language-specific, and depending on the specific features of that language.

There are different languages for writing smart contracts in Ethereum, among which, in this paper we just focus on the Solidity programming language. Solidity is very similar to  JavaScript, but it includes more functions to support writing smart contracts. To design effective mutation operators for Solidity smart contracts, 
we first conducted an extensive study on the known real world bugs in the scope of smart contracts. Our sources of investigation are academic papers, open-source smart contracts repositories including  open or closed issues on Github,  articles in the websites like Medium and Q\&A websites like Stack-Exchange~\cite{exchange}. We ended up with a collection of most repeated bugs that may happen in the implementation of a smart contract in Solidity programming language. Studying the bugs in smart contracts, we categorize them to two groups:
\begin{enumerate}
	\item {\em Classic Bugs:} These bugs occur in almost any programming language, from which we can mention arithmetic issues or logical bugs (inside conditions).
	\item {\em Solidity Bugs:} These faults are mostly related to the Solidity programming languages, and the distributed nature of blockchain and smart contracts.
\end{enumerate}

Hence, we conclude that classical mutation operators designed for general-purpose programming languages, e.g. JavaScript, are not sufficient for the Ethereum platform, and we need to design other mutation operators to mimic the Solidity specific bugs. So, we divide our mutation operators into two groups: \textit{(i)} Classic Mutation Operators, and \textit{(ii)} Solidity Mutation Operators. 


\subsection{Classic Mutation Operators (CMO)}\label{ssec:Designing of CMO}
Classic mutation operators include a set of mutation operators, which can be used  for almost any programming languages (with maybe some minor differences due to the difference in syntax). Among these operators, we can mention arithmetic and logical operators.
They are designed to target a group of bugs that are common in most programming languages.
 Effective classic mutation operators have been designed for different languages, which also can be used for Solidity smart contracts~\cite{ma2006evaluation}.
These operators mostly contain insertion, deletion, and replacement of arithmetic and logical operators, some of which are depicted in Table~\ref{table:classic mutation operators}.
\begin{table}[]
	\centering
	\begin{tabular}{|l|l|}
		\hline
		Operators & Description \\ \hline
		ABS          &     ABSolute value insertion                     \\ \hline
		AOR       &         Arithmetic Operator Replacement                 \\ \hline
		CRP      &             Constants RePlacement             \\ \hline
		CRR    &          Constants for Reference Replacement                \\ \hline
		LCR      &         Logical Connector Replacement                 \\ \hline
		ROR     &          Relational Operator Replacement                \\ \hline
		RCR     &           Reference for Constant Replacement               \\ \hline
		FDL       &          Formula DeLetion                \\ \hline
		FRC       &         Formula Replacement with Constant                 \\ \hline
		RFR       &          Reference Replacement                \\ \hline
		UOI    &          Unary Operator Insertion               \\ \hline
		SCL    & Swap Code Lines \\ \hline
	\end{tabular}
	\caption{Classic mutation operators}
	\label{table:classic mutation operators}
\end{table}
\subsection{Solidity Mutation Operators (SMO)}\label{ssec:Designing of SMO}
Considering our list of known real-world bugs, we have designed 9 classes of Solidity-specific mutation operators to mimic the common security and software bugs in smart contracts. In this section, we discuss different classes of Solidity-specific bugs and our designed mutation operators for them.
\subsubsection{Overflow-Underflow}\label{sssub:Overflow-Underflow}
Overflow and underflow bugs are common in many programming languages, but in Solidity smart contracts, an underflow or overflow breach can cause huge monetary losses.
Overflow happens when a number exceeds the upper bound of its type. For example, assume a \lstinline $uint32$ variable that has the maximum value of $2^{32}-1$. If this variable increases in an operation beyond its maximum value, it will return to its minimum value of $0$. More specifically, consider this scenario of a smart contract, where it is needed to increase someone's balance, but because of overflow, the balance value will be set to $0$. 
A developer should recheck each statement using some guard functions, to avoid overflow or underflow related bugs. In Listing~\ref{lst:overflow}, you can see an example for a smart contract containing an overflow bug, and a method to secure it.
\begin{lstlisting}[caption={Overflow bug},label={lst:overflow},numbers=none]
mapping (address => uint32) public balanceOf;

// INSECURE
function transfer(address to, uint32 value) {
	//Check if sender has balance
	require(balanceOf[msg.sender] >= value);
	//Add and subtract new balances
	balanceOf[msg.sender] -= value;
	balanceOf[to] += value;
}

// SECURE
function transfer(address to, uint32 value) {
	//Check if sender has balance and for overflows */
	require(
		balanceOf[msg.sender] >= value 
		&& 
		balanceOf[to] + value >= balanceOf[to]
	);
	
	//Add and subtract new balances 
	balanceOf[msg.sender] -= value;
	balanceOf[to] += value;
}
\end{lstlisting}
Similar bug happens with underflow, when one tries to decrease a \lstinline|uint32| variable, for example, below its minimum value (i.e. $0$), and instead of getting a negative value, the variable gets its maximum value ($2^{32}-1$). For example, consider someone tries to withdraw all of her balance from a contract, but instead of setting her balance to $0$, she ends up with the maximum value of the variable type, because of a simple underflow. Listing~\ref{lst:underflow} shows an example source code containing an underflow bug:
\begin{lstlisting}[caption={Underflow bug},label={lst:underflow},numbers=none]
mapping (address => uint32) public balanceOf;

// INSECURE
function transfer(address to, uint32 value) {
	//Check if sender has balance
	require(balanceOf[msg.sender] >= value);
	//Add and subtract new balances
	balanceOf[msg.sender] -= value;
	balanceOf[_to] += value;
}

// SECURE
function transfer(address to, uint32 value) {
	//Check if sender has balance and for overflows
	require(
		balanceOf[msg.sender] >= value 
		&& 
		balanceOf[to] - value <= balanceOf[to]
	);
	
	//Add and subtract new balances
	balanceOf[msg.sender] -= value;
	balanceOf[to] += value;
}
\end{lstlisting}
In Listings~\ref{lst:overflow} and~\ref{lst:underflow}, assuming that balances are real assets like \textit{Ether}, simple overflow or underflow can cost millions of dollars. 
To detect these bugs, we propose Overflow-Underflow mutation operators, which are depicted in Table~\ref{table:overflow-underflow}. In IST and DST operators, the size of a type is increased/decreased respectively. For example, the type \lstinline $int32$ is replaced by \lstinline $int8$, or \lstinline $int64$.

\begin{table}[]
	\centering
	\begin{tabular}{|l|l|}
		\hline
		Operators & Description \\ \hline
		USP          &       Unsigned to Signed Replacement                     \\ \hline
		PSU       &           Signed to Unsigned Replacement                 \\ \hline
		IST      &               Increase Size of a Type             \\ \hline
		DST    &           Decrease Size of a Type                \\ \hline
	\end{tabular}
	\caption{Mutation operators for Overflow-Underflow}
	\label{table:overflow-underflow}
\end{table}

\subsubsection{Access Control}\label{sssec:Acess Control}
It is a must for every developer to understand how she can restrict parts of her code from other parts. In smart contracts, the access controls is very important and the corresponding bugs can be very dangerous and potentially leads to big loss of assets. 
In Solidity, there are four levels of access control, namely Public, External, Private, and Internal~\cite{soldoc}. The difference between these types can be seen in Table~\ref{table:AClevels}.
\begin{table}[]
	\centering
	\begin{tabular}{|l|p{5.7cm}|}
		\hline
		Access Control & Definition \\		\hline
Public & All can access  \\ \hline
	External & Cannot be accessed internally, only externally  \\ \hline
	Internal & Only this contract and contracts deriving from it can access  \\ \hline
	Private & Can be accessed only from this contract  \\ \hline
	\end{tabular}
\caption{Access control levels in Solidity}
\label{table:AClevels}
\end{table}

Listing~\ref{lst:access control} shows an example for access control bug in a smart contract. This contract has a public function called \lstinline {helpCharity} that everyone can call to transfer some assets to the contract address from her balance. Also, there is another function named \lstinline $transfer$, which should be called from  \lstinline {helpCharity} to transfer asset to the charity's address. As this function is mistakenly defined  as public, an attacker can call \lstinline $transfer$ function and transfer all assets to her address, instead of the charity's address.
\begin{lstlisting}[caption={Access control bug},label={lst:access control},numbers=none]

mapping (address => uint256) public balanceOf;
//Public function so everyone can help the charity
function helpCharity(uint256 value) public {
	let charityAddress = 0xe0f5206bbd039e7b0592d8918820024e2a7437b9;
	transfer(to: charityAddress, value: 1);
	balanceOf[msg.sender] -= value;
}
// transfer mistakenly defined as a public instead of private
function transfer(address to, uint256 value) public {
	balanceOf[to] += value;
}
\end{lstlisting}
To detect the access control bugs, we have designed a set of mutation operators, which are listed in Table~\ref{table:access control}.
\begin{table}[]
	\centering
	\begin{tabular}{|l|l|}
		\hline
		Operators & Description \\ \hline
		PuPrR          &         Public to Private Replacement                     \\ \hline
		PuIR       &            Public to Internal Replacement                \\ \hline
		PuER      &                 Public to External Replacement            \\ \hline
		PrPuR    &             Private to Public Replacement                \\ \hline
		PrIR          &         Private to Internal Replacement                   \\ \hline
		PrER       &             Private to External Replacement                 \\ \hline
		IPuR      &                 Internal to Public Replacement             \\ \hline
		IPrR    &             Internal to Private Replacement                \\ \hline
		IER          &         Internal to External Replacement                     \\ \hline
		EPuR       &             External to Public Replacement                \\ \hline
		EPrR      &                 External to Private Replacement             \\ \hline
		EIR    &             External to Internal Replacement               \\ \hline
	\end{tabular}
	\caption{Mutation operators for access control}
	\label{table:access control}
\end{table}
\subsubsection{Transaction Call Mechanism}\label{sssec: Transaction Call Mechanism}
In Ethereum, we can use transactions to send an asset from one account to another account or trigger another contract's code. 
There are three methods for creating a transaction in Solidity smart contracts. These methods are different in terms of gas usage and their return value, and hence, each one should be used in appropriate situations. These three methods are as follows:

\begin{enumerate}
	\item \lstinline $send$: It only includes 21000 \textit{wei} (the smallest denomination of \textit{Ether}) into the transaction. That means the execution process of the called function must not exceed 21000 \textit{wei}, otherwise the transaction will be incomplete, and the \lstinline|send| returns  \lstinline|false|.
	\item \lstinline $transfer$: \lstinline $transfer$ also includes only 21000 \textit{wei} into the transaction. However, if the execution of a function exceeds 21000 \textit{wei} of gas, it will throw an exception, rather than returning \lstinline|false|.
	\item \lstinline $call$: Using this method can be dangerous, as it transfers all the remaining gas for the transaction to the called function.
\end{enumerate}

An example of a bug caused by inappropriate usage of transaction call is {\em reentrancy}. This attack occurs when a function from the first smart contract makes an external call to another one using \lstinline $call$ function.
Listing~\ref{lst:reentrancy example} shows an example  of this bug, where an attacker can deploy a contract and define its fallback function to rerun the \lstinline|withdraw| function. 
After deploying the contract, she can call \lstinline $Victim$'s \lstinline|withdraw| function, and \lstinline $msg.sender.call.value(amount)()$ is going to send some assets to \lstinline $Attacker$'s contract, which results in calling the fallback function of the \lstinline|Attacker| that will call the \lstinline|withdraw| function again, and so on. Since \lstinline $Victim$ uses \lstinline $call$ as a mechanism to send asset, \lstinline $call$ is going to send all the remaining gas to the attacker's contract and the attacker can take control of the transaction flow and recursively call \lstinline $withdraw$ function, till all the gas burns. Therefore, in this scenario an attacker can steal all the contract's assets before the \lstinline|Victim| contract sets her balance to zero. But, if \lstinline $Victim$ uses \lstinline $send$ function instead of \lstinline $call$, it would only use 21000 \textit{wei}, and \lstinline $Attacker$'s fallback function would throw ``OutofGas" exception because of lack of gas.
\begin{lstlisting}[caption={Reentrancy Attack},label={lst:reentrancy example},numbers=none]
//Victim function
function withdraw() external {
	uint256 amount = balances[msg.sender];
	require(msg.sender.call.value(amount)());
	balances[msg.sender] = 0;
}

//Attacker fallback function
function() external payable {
	Victim v;
	v.withdraw();
}
\end{lstlisting}
We have designed 6 mutation operators to take care of transaction call mechanism bugs that can be seen in Table~\ref{table:transaction call mechanism mutant operators}.
\begin{table}[]
	\centering
	\begin{tabular}{|l|l|}
		\hline
		Operators & Description \\ \hline
		TCR          &        Transfer to Call Replacement                    \\ \hline
		TSR       &            Transfer to Send Replacement                \\ \hline
		SCR      &                 Send to Call Replacement            \\ \hline
		STR    &            Send to Transfer Replacement                \\ \hline
		CSR          &         Call to Send Replacement                   \\ \hline
		CTR       &             Call to Transfer Replacement                \\ \hline
	\end{tabular}
	\caption{Mutation operators for transaction call mechanism}
	\label{table:transaction call mechanism mutant operators}
\end{table}
\subsubsection{Guard Mechanism}\label{sssub:Guard Mechanism}
In Solidity, it is possible to use {\em guards} as a way of handling errors. Guards are some state-reverting exceptions which will undo all changes in the EVM if a condition (a Boolean expression) is not met.

There are three guard functions available in Solidity programming language, which are \lstinline $assert$, \lstinline $require$, and \lstinline{revert}. Each one has its own behavior, and a developer should be careful about using the appropriate one. Careless usage can result in freezing the contract, or making it vulnerable to attacks.
When we use \lstinline $assert$ as a guard function, this operator will use all the associated gas, but \lstinline $require$ and \lstinline $revert$ will send back the remaining gas.
As an example, in Listing~\ref{lst:guard mechanism}, if the developer forgets to use \lstinline $require$ in the \lstinline $justOwner$ function, anyone can call \lstinline $justOwner$ which will lead to abnormal and unexpected behavior. 
\begin{lstlisting}[caption={Guard mechanism bug},label={lst:guard mechanism},numbers=none]
address owner;
function justOwner() {
	require(msg.sender == owner)
	//if someone rather than owner call this function something bad will happen.
}
\end{lstlisting}
Table~\ref{table:guard mechanism mutant operators} shows our mutation operators for the guard mechanism.
\begin{table}[]
	\centering
	\begin{tabular}{|l|l|}
		\hline
		Operators & Description \\ \hline
		ARevR          &        Assert to Revert Replacement                    \\ \hline
		AReqR       &            Assert to Require Replacement                \\ \hline
		RevReqR      &                 Revert to Require Replacement           \\ \hline
		RevAR    &             Revert to Assert Replacement                \\ \hline
		ReqAR          &         Require to Assert Replacement                  \\ \hline
		ReqRevR       &             Require to Revert Replacement                \\ \hline
		AReq      &                 Add Require             \\ \hline
		AA    &             Add Assert                \\ \hline
		ARev          &         Add Revert\\ \hline
		DReq       &             Delete Require                \\ \hline
		DRev      &                 Delete Revert             \\ \hline
		DA    &             Delete Assert              \\ \hline
	\end{tabular}
	\caption{Mutation operators for guard mechanism}
	\label{table:guard mechanism mutant operators}
\end{table}
\subsubsection{Transaction  Origin}\label{sssub: Tx Origin Attacks}
There are a number of fields in the \lstinline|msg| object that provide more information for the receiver of the message (transaction). For example, \lstinline|msg.sender| identifies who has triggered the transaction, \lstinline|msg.value| contains the value of attached assets, and \lstinline|msg.data| returns the attached date to the transaction, if any. 
Also,  \lstinline|tx.origin| can be used to find out the original first user who has triggered the transaction.  A transaction can be the result of a chain of other transactions, which have called each other, and in these cases,  \lstinline|tx.origin| can identify who is the root of this call chain.


Inappropriate usage of \lstinline $tx.origin$ and \lstinline $msg.sender$ can put a smart contract into the danger of an attack. As an example, in Listing~\ref{lst:tx origin attacks}, there are two smart contracts, called \lstinline $Victim$ and \lstinline $Attacker$. In \lstinline $Victim$, only the owner of the smart contract can call the \lstinline $transfer$ function and send assets to an address. The developer has mistakenly used \lstinline $tx.origin$ to see whether the caller is the owner or not. If the owner wants to transfer some assets to the \lstinline $Attacker$ contact, after triggering the transaction, assets will be transferred to the \lstinline $Attacker$ contract, resulting in running its fallback function. The statements in the fallback function causes to transfer any number of funds to the \lstinline $Attacker$, because the very first origin of the  transaction is the \lstinline|Victim|'s owner. 
To prevent this, the developer should check \lstinline|msg.sender|, instead of \lstinline|tx.origin|.
\begin{lstlisting}[caption={A transaction origin atack},label={lst:tx origin attacks},numbers=none]
//VICTIM
contract Victim {
	address owner;
	constructor() public {
		owner = msg.sender;
	}
	function transfer(address to, uint64 amount)
	 public {
		require(tx.origin == owner);
		to.call.value(amount)();
	}
}

//ATTACKER
interface Victim {
	function transfer(address to, uint amount);
}
contract Attacker {
	address owner;
	constructor() public {
		owner = msg.sender;
	}
	function() payable public {
		let randomValue = 64;
		Victim(msg.sender).
		transfer(owner, randomValue);
	}
}
\end{lstlisting}
Table~\ref{table:Tx origin mutation operators} shows our mutation operators, designed for Transaction Origin attacks.
\begin{table}[]
	\centering
	\begin{tabular}{|l|l|}
		\hline
		Operators & Description \\ \hline
		TMR          &         Tx.origin to Msg.sender Replacement                   \\ \hline
		MTR          &         Msg.sender to Tx.origin Replacement                  \\ \hline
	\end{tabular}
		\caption{Mutation operators for Transaction Origin}
	\label{table:Tx origin mutation operators}
\end{table}
\subsubsection{\lstinline $selfdestruct$ Operator}\label{sssub: Manipulate selfdestruct}
The \lstinline|selfdestruct| function (previously known as  \lstinline $suicide$), helps developers to destruct a smart contract. 
This function takes an address as argument to transfer all the stored assets in the smart contract to that, before destructing the smart contract. For example, calling \lstinline $selfdestruct(addr)$ sends all the contract's current balance to the address \lstinline $addr$.

This operator is useful, when we are done with a smart contract, as it transfers assets to a given address with consuming less gas amount than simple transfer using \lstinline $addr.send(this.balance)$.
In fact, because of freeing up some space in the EVM,  \lstinline|selfdestruct| actually uses negative gas.
Thus, using this operator wisely can be helpful, however,  if a developer mistakenly uses this function, it will lead to destroying the smart contract and it may cause loss of stored assets. In order to deal with this bug,  we have designed two mutation operators to remove \lstinline $selfdestruct$ from the code, and swap its location to adjacent lines (Table~\ref{table:Manipulate selfdestruct}).
\\
\begin{table}[]
	\centering
	\begin{tabular}{|l|l|}
		\hline
		Operators & Description \\ \hline
		RSF          &          Remove   selfdestruct from a Function \\ \hline
		SSL          &         Swap selfdestruct's location to adjacent lines \\ \hline
	\end{tabular}
	\caption{Mutation operators for \lstinline $selfdestruct$}
	\label{table:Manipulate selfdestruct}
\end{table}
\\
\subsubsection{Constant Properties}\label{sssub:Manipulate constant properties}
Each smart contract can use many constant values, including smart contract or user addresses, number of tokens (e.g., the number of tokens to issue), gas amount on each transaction call, etc.
Setting wrong constant values can lead to incorrect behavior of a smart contract. For example, using wrong constant address can cause sending money to a wrong user, or running wrong function from a malicious contract. Settings wrong gas amounts may fail a transaction, or wrong value of issued token can cause huge monetary losses.
Therefore, developers should set the constant values carefully in smart contracts before deploying them on EVM.
Table~\ref{table: manipulate constants properties mutant operators} depicts our mutation operator designed for constants values used in a smart contract.
\begin{table}[]
	\centering
	\begin{tabular}{|l|l|}
		\hline
		Operators & Description \\ \hline
		CAA          &          Change Address to another Address                \\ \hline
				CDG          &          Increase and Decrease in Gas amount                   \\ \hline
						CCV &          Change in any Constant Value                   \\ \hline
	\end{tabular}
	\caption{Mutation operators for manipulating constants}
	\label{table: manipulate constants properties mutant operators}
\end{table}
\subsubsection{Function Modifiers}\label{sssub: Manipulate Modifiers}
Function modifiers can be defined as Boolean conditions added to a function declaration. When a function with modifier is called, first the modifiers are validated, and only if all its modifiers are satisfied, the function begins to execute. 
The function modifiers are widely used to check the authority of Ethereum users for calling a smart contract's function. Hence, wrong or buggy modifier can cause unauthorized access to important functions of a smart contract.
As an example, in Listing~\ref{lst:manipulate modifiers}, there is a function, called \lstinline $justOwner$ with the  \lstinline $onlyOwner$ modifier. This modifier restricts calling this function to the contract owner. Using this modifier can be very important, as the function may include transferring the contract assets. You can see that a small bug in the  modifier, such as changing \lstinline $==$ to \lstinline $!=$, can cause unauthorized access of an attacker.
\begin{lstlisting}[caption={An example of using modifier},label={lst:manipulate modifiers},numbers=none]
address owner;

modifier onlyOwner {
	require(msg.sender == owner);
}

function justOwner() onlyOwner {
	// transferring the contract assets
}
\end{lstlisting}
Table~\ref{table:manipulate modifiers mutant operators} shows our proposed mutation operators for manipulating modifiers. These mutation operators change each modifier to the constant values of \lstinline|true| or \lstinline|false|. Selected test cases by these operators can better check the correctness of the implemented modifiers.
\begin{table}[]
	\centering
	\begin{tabular}{|l|l|}
		\hline
		Operators & Description \\ \hline
		CMT          &          Change Modifier to \lstinline|true|                   \\ \hline
		CMF       &              Change Modifier to \lstinline|false|                \\ \hline
	\end{tabular}
	\caption{Mutation operators for manipulating modifiers}
	\label{table:manipulate modifiers mutant operators}
\end{table}
\subsubsection{Constructor Name}\label{sssub: Manipulate Smart Contract's Constructor}
In Solidity, a smart contract can be initialized by a constructor, where the required initialization statements of a smart contract take place before deploying it to the EVM. 
For example, initialization of the owner property of the smart contract can be performed in its constructor. Correct spelling of the constructor name is very important, as a mistake in its spelling will cause Solidity not recognizing the function as a constructor. On the other hand, if a function is mistakenly named as the constructor's name, it is going to be behaved like a constructor, and can cause unexpected behaviors.
Listing~\ref{lst:Constructor} is an example, where misspelling of a constructor function can cause damage to the contract. In this example, a typo in the name of the constructor made it a simple regular function named \lstinline $Exampl$. Thus, any user can call this method and introduce herself as the contract owner.
\begin{lstlisting}[caption={An example of a bug caused by constructor name},label={lst:Constructor},numbers=none]
address owner;

contract Example {
	// Correct Behaviour
	function Example(address add) public {
		owner = add
	}


	// Wrong Behaviour (typo on contract name)
	function Exampl(address add) public {
		owner = add
	}
}
\end{lstlisting}
Table~\ref{table:Manipulate Smart Contract's Constructor Name} shows our mutation operators for manipulating the smart contract's constructor name.
\begin{table}[]
	\centering
	\begin{tabular}{|l|l|}
		\hline
		Operators & Description \\ \hline
		CCN          &          Change Constructor Name to something else                   \\ \hline
		CFC       &              Change a Function's name  to Constructor                \\ \hline
	\end{tabular}
	\caption{Mutation operators for manipulating constructor name}
	\label{table:Manipulate Smart Contract's Constructor Name}
\end{table}

\section{Evaluation}\label{Evaluation}

Mutation testing is a well-known technique in software testing, and different mutation operators have been proposed for several languages. There are a few methods to evaluate the effectiveness of the designed mutation operators~\cite{jia2010analysis,gopinath2014mutations}. One way is to analyze the real faults already detected in software developed in the studied language, and find the number of faulty codes that can be regenerated by applying one of the mutation operators to the corresponding corrected code. High percentage in this evaluation can indicate the resemblance of the mutation operators to the real faults, and can be an indication to the power of the mutants in selecting effective test cases (the ones that can find the bugs in the code).

To evaluate the effectiveness of our proposed mutation operators using this technique, we first gathered a comprehensive set of real world known bugs from a number of available lists including~\cite{atzei2016survey, not-so-smart-contracts,dasp,awesome-buggy-erc20,solidity-known-attacks, luu2016making}. Then, we applied our mutation operators to the source code of the fixed version of these smart contracts to see if we can generate the previously detected bugs. Table~\ref{table:evaluation} summarizes our evaluation results for these real world case studies, which shows that our mutation operators can regenerate 10 out of 15 real world bugs. Here, we list all the designed mutation operators related to each real world bug, and the operators that could regenerate the bug have been marked with ``*''.

\begin{table*}
	\centering
	\scriptsize
	\begin{tabular}{|p{2cm}|l|p{2.4cm}|l|l|c|}
		\hline
		\textbf{Bug Class} & \textbf{The Bug} & \textbf{Related Operators} & \textbf{Value} & \textbf{Date} & \textbf{Reproducable?} \\\hline
		Re-Entrancy & The DAO \cite{dao-report-slockit} & SCL*, SCR & 3M \textit{Ether} ($\sim$ \$50M) & June 2016 & \OK \\\hline
		 \multirow{2}{2cm}{Delegated Call} & Parity: ``MultiSig" \cite{parity-multisig-bug} & IPuR*, IPrR, IER & 153K \textit{Ether} ($>$ \$30M) & July 2017 & \OK \\
		 & Parity: ``I Accidentally Killed It!" \cite{parity-accidentally-kill} & RSF, ASF* & 500K \textit{Ether} ($\sim$ \$300M) & Nov 2017 & \OK \\\hline 
		 \multirow{3}{2cm}{Arithmetic} & PoWN Overflow \cite{powh-underflow} & IST, DST, DReq*  & 886 \textit{Ether} ($\sim$ \$800K) & Feb 2018 & \OK \\
		 & BeautyChain Batch Overflow Bug \cite{bec-batch-overflow}& IST, DST, DA* & $10^{58}$ \textit{BEC} & April 2018 & \OK \\
		 & ICON Disable All Transactions \cite{icon-fatal} & LCR* & Potentially \$800M & Jun 2016 & \OK \\\hline
		 \multirow{2}{2cm}{Constructor Name} & Rubixi \cite{rubixi} & CCN*, CFC & Potentially 108 \textit{Ether} & April 2016 & \OK \\
		 & Morph Case Sensitive Constructor \cite{morph-token-constructor} & CCN*, CFC & $\sim$ 400 \textit{Ether} & June 2018 & \OK \\\hline
		 \multirow{2}{2cm}{Timestamp-dependence} & The Run \cite{why-smart-contracts-fail} & & & & \NOK \\
		 & GovernMental: ``Timestamp" \cite{atzei2016survey} & & & & \NOK \\\hline
		  \multirow{3}{2cm}{Not Checked Return Values} & KoET Return Check \cite{koet-postmortem} & CSR*, SCR, DReq* & 98.5 \textit{Ether} & Feb 2016 & \OK \\
		  & Etherpot Blockhash Bug \cite{etherbot-blockhash} & & & Sept 2016 & \NOK \\
		  & SmartBillions Randomness \cite{smart-billions} & & $\sim$ 400 \textit{Ether} & Oct 2017 & \NOK \\\hline
		  Dynamic Libraries & Re-Entrancy Honey Pot \cite{reentrancy-honeypot} & CAA* & 1 \textit{Ether} & Feb 2018 & \OK \\\hline
		  Denial of Service & Governmental: ``Too Much Gas" \cite{governmental-too-much-gas} &  & 1100 \textit{Ether} & Apr 2016 & \NOK \\\hline 
	\end{tabular}
	\caption{Evaluation Results}\label{table:evaluation}
\end{table*}

In the following, we discuss more details on some of these faults in the real world examples, especially the ones that cannot be regenerated by our mutation operators. 
``Re-Entrancy'' and ``Delegated Call'' bugs have been the most costly faults in the Ethereum history. Our mutation operators can successfully regenerate these classes of faults in smart contracts. Arithmetic issues are also the other most repeated faults in smart contracts. To detect these arithmetic issues, we have proposed  Solidity-specific mutation operators, in addition to the classical operators for arithmetic. These operators help us to cover more real faults in the arithmetic class of bugs in Solidity language.


Another class of known real world bugs are Timestamp-dependence faults, which are very common in lottery ERC20 tokens (e.g. \cite{atzei2016survey, why-smart-contracts-fail}). This bug happens when a  smart contracts uses the block timestamp values as a random seed, ignoring the fact that a malicious miner can manipulate the timestamp value within 900 seconds to her benefits~\cite{luu2016making}. Similarly, some smart contracts use the \lstinline|blockhash| functions to compute the hash values of blocks. This function can only compute the hash values for the last 256 Ethereum blocks and in other cases, it will return zero. There are many faulty lottery smart contracts that use \lstinline|blockhash| function to generate random numbers~\cite{etherbot-blockhash, smart-billions}. To avoid these two faults, we suggest developers to use 3rd party libraries, which provide secure random generators~\cite{chatterjee2019probabilistic, rando}. Since the solutions (corrections) for the bugs related to generating random numbers are complicated mechanisms, we could not design any mutation operator that can regenerate this bug from the corrected code.

To evaluate our mutation operators in practice, we have also extended a well-know tool for generating mutants, called \textit{Universal Mutator}~\cite{universal}, and added our Solidity-related mutation operators to it. The enhanced tool accepts a Solidity smart contract and generates all its possible mutants based on our proposed operators. One of the great features about this tool is detecting invalid (the ones that cannot be compiled) and redundant mutants. 
The source code for this tool is available online on Github~\cite{universal-mutator}, and can be used by other researchers and smart contract developers. We used this tool to generate the mutants for a number of well-known smart contracts. The number of valid, invalid, and redundant mutants generated for each smart contract can be seen in
 Table~\ref{table:Mutants generated by Universal Mutator}.
\begin{table}[]
	\centering
	\begin{tabular}{|l|l|l|l|l|}
		\hline
		Smart Contract & Valid & Invalid & Redundant & Total \\ \hline
		Smartex          &     60 & 57 & 25 & 142                     \\ \hline
		ethBank       &         154 & 22 & 47 & 223                \\ \hline
		NEST\_LoanContract      &             129 & 169 & 89 & 387             \\ \hline
		BITNOMO    &          71 & 136 & 65 & 272                \\ \hline
		WOR      &         45 & 83 & 42 & 170                \\ \hline
	\end{tabular}
	\caption{Mutants generated by the extended universal mutator}
	\label{table:Mutants generated by Universal Mutator}
\end{table}

\section{Related Work}\label{ssec:Related Work}
In this section, we discuss the related works in two areas of analyzing  smart contracts and mutation testing.
\subsection{Mutation Testing}\label{ssec:Mutation Testing in Software Engineering}
Mutation Testing was first proposed by Lipton in~\cite{lipton1971fault}.
The technique can be used in different levels of testing, including unit, integration, and specification. 
Mutation can be used to  identify effective  test suite that developers can use to test their application~\cite{geist1992estimation}. Mutation operators are often designed to mimic the common faults in a programming language. 
The technique has been widely used for test design in different  programming languages, including Fortan~\cite{budd1979mutation}, Ada~\cite{bowser1988reference}, C~\cite{agrawal1989design}, and Java~\cite{epstein1963physiological}. It has even been used to test non-functional properties of a software, such as energy testing of Android applications~\cite{jabbarvand2017mudroid}.

\subsection{Analysis of Smart Contracts}\label{ssec:Mutation Testing in Smart Contracts}
There are several works on analyzing  smart contracts for bugs and vulnerabilities. They can be categorized into three groups; static analysis techniques, formal verification, and testing. Luu et al. in~\cite{luu2016making} introduced OYENTE, which is a tool to statically analyze smart contracts using symbolic execution . Symbolic execution can be used to identify the inputs  needed to reach to specific parts of a code. Bhargavan et al. in~\cite{bhargavan2016formal} introduced a framework to convert the bytecode of a smart contract to $F^*$, a functional programming language, and then verify a set of specifications for it. The limitation of this framework is due to the complexity of verification, and as reported in the paper, only 49 smart contracts out of a  list of 396 contracts could be verified by this framework. Kalra et al. in~\cite{kalra2018zeus} introduced Zeus that uses abstract interpretation and symbolic model checking to formally check smart contracts. Zeus has less false positives compared to similar frameworks. However, it does not support some Solidity operators, like \lstinline $selfdestruct$ or \lstinline $throw$. Brent et al. in~\cite{brent2018vandal} introduced Vandal as a security analysis tool for Ethereum smart contracts. Vandal uses logic-driven program analysis to analyze smart contracts and detect the security bugs and vulnerabilities.
Finally, Jiang et al. developed ContractFuzzer which is a fuzzer to test smart contracts for vulnerabilities and security drawbacks~\cite{jiang2018contractfuzzer}.

There is one recent work on testing smart contracts using  mutation testing approach~\cite{wu2019mutation}. In this work, Wu et al. proposed 15 mutation operators based on Solidity documentations. In comparison, we have surveyed a comprehensive list of reported  real bugs in Ethereum smart contracts, and designed our mutation operators based on these bugs. We have also included a set of mutation operators for Solidity-specific features. Using this approach, we have  a more complete list of operators, from which we can mention the operators for  overflow-underflow, transaction call mechanisms, \lstinline|selfdestruct|, and modifiers. We have also suggested including more classical operators such as SCL, ABS, etc to better detect possible bugs in Solidity smart contracts. Considering the evaluation, we have evaluated our mutation operators by checking their power in regenerating the real faults. We have also extended the \textit{Universal Mutator} with our rules to automatically generate mutants for Solidity smart contracts.

\section{Conclusion}\label{Conclusion}
Smart contracts are immutable, verifiable, and autonomous software, hosted and ran on blockchains, like Ethereum. Due to their immutable nature, smart contracts cannot be changed, once they are deployed on the blockchain. Therefore, it is important for developers to fully analyze their smart contracts before deployment. 
In this research, we analyzed a comprehensive list of known bugs  in Solidity smart contracts, and proposed 10 classes of mutation operators. Our experiments show that our  operators can regenerate the real  faults for 10 smart contracts out of 15 famous buggy ones. We have also extended the universal mutator tool with our mutation operators to automatically generate mutants for smart contracts. We believe that our designed operators can help Solidity developers (and testers) to better develop bug-free and safe smart contracts.

\bibliography{mybibfile}

\end{document}